\documentclass[aps,prl,twocolumn,showpacs,superscriptaddress,groupedaddress]{revtex4}  

\usepackage{graphicx}
\usepackage{dcolumn}
\usepackage{bm}

\usepackage{flushend}

\usepackage{amsmath,amssymb}
\usepackage{mathtools}
\usepackage{hyperref}

\newcommand{\LQ}{\Lambda_{\rm QCD}}

\newcommand{\bea}{\begin{eqnarray}}
\newcommand{\eea}{\end{eqnarray}}
\newcommand{\simgt}{\hbox{ \raise3pt\hbox to 0pt{$>$}\raise-3pt\hbox{$\sim$} }}
\newcommand{\simlt}{\hbox{ \raise3pt\hbox to 0pt{$<$}\raise-3pt\hbox{$\sim$} }}

\newcommand{\be}{\begin{equation}}
\newcommand{\ee}{\end{equation}}

\begin{document}

\begin{flushright}
TU--1236, YITP--24--75
\end{flushright}

\title{Two-loop Quarkonium Hamiltonian in Non-annihilation Channel}

\author{Go Mishima}
\author{Yukinari Sumino}
 \affiliation{Department of Physics, Tohoku University,
Sendai, 980--8578 Japan
}
\author{Hiromasa Takaura}
\affiliation{%
Center for Gravitational Physics and Quantum Information, Yukawa Institute for Theoretical
 Physics, Kyoto University, Kyoto 606-8502, Japan
}%


\date{\today}

\begin{abstract}
We calculate the two-loop heavy quarkonium Hamiltonian within potential-NRQCD effective field theory in the non-annihilation channel. 
This calculation represents the first non-trivial step towards determining the N$^4$LO Hamiltonian
in the weak coupling regime.
The  large amount of computation is systematically handled by employing the $\beta$ expansion, differential equations for master integrals, and adopting a single-step matching procedure, in contrast to the conventional two-step approach.
\end{abstract}

\maketitle



The heavy quarkonium system provides an ideal laboratory for investigating theoretical and phenomenological aspects of QCD in depth \cite{QuarkoniumWorkingGroup:2004kpm,Brambilla:2010cs,Sumino:2014qpa}.
In particular, analyses of quarkonium systems in the weak coupling regime have enabled precise examinations 
of QCD, in
the case that their relevant physical scales are much larger than the typical QCD
scale $\LQ \sim 300$~MeV.
For instance, these analyses have led to the determinations of fundamental physical constants such as the charm quark mass, 
bottom quark mass, 
and strong coupling constant $\alpha_s$ \cite{Kiyo:2015ufa,Peset:2018ria}.
These analyses have provided valuable constraints on Grand Unified Theories (GUTs), specifically through the precise bottom-tau mass ratio.
Moreover, investigations of the top-antitop quark threshold region in future $e^+e^-$ collider experiments are expected to be a major focus of high-energy physics, driving considerable theoretical efforts towards high-precision predictions \cite{Beneke:2013jia,Vos:2016til}.

The quarkonium Hamiltonian\footnote{
The quarkonium Hamiltonian is given by a set of quantum mechanical operators which act on the color-singlet quark-antiquark composite field in the pNRQCD Lagrangian.
} 
within the potential Nonrelativistic QCD (pNRQCD) effective field theory (EFT) \cite{Pineda:1997bj,Brambilla:2004jw}
has played
a principal role in these precision analyses.
The Hamiltonian has been known up to the 
next-to-next-to-next-to-leading order (N$^3$LO) accuracy for
more than a decade \cite{Kniehl:2002br,Anzai:2009tm,Smirnov:2009fh}, and 
it took roughly a decade to calculate the N$^3$LO Hamiltonian after the N$^2$LO Hamiltonian
was completed \cite{Schroder:1999sg}.

In this paper, as a first non-trivial partial calculation of the N$^4$LO quarkonium Hamiltonian,
we calculate the Hamiltonian at the two-loop level, or at ${\cal O}(\alpha_s^3)$,
and up to ${\cal O}(\beta^0)$ [relative ${\cal O}(\alpha_s^2\beta^2)$ compared to the LO $\sim \alpha_s/\beta^2$
contribution].
Here, $\beta$ denotes the velocity of the heavy quark or antiquark 
in the center-of-mass (c.m.)\ frame and is a small expansion parameter of the quarkonium system.
To reduce the labor of the calculation, we compute only the contributions from the non-annihilation channel
of the quark and antiquark.
Our motivation is to establish a calculational procedure for such a complex calculation.

We consider the ${\rm SU}(3)$ color gauge
theory with $n_h$ heavy quark flavors (each with the same mass $m$) and $n_l$ massless quark
flavors.
(For convenience we call this theory as QCD.)
We calculate the scattering amplitude between a heavy quark $Q$ and a heavy antiquark $\bar{Q}'$ of
different flavors (with mass $m$).
In this case only diagrams in the non-annihilation channel contribute.
We evaluate this amplitude in the $\beta$ expansion ($1/m$ expansion) in the c.m.\ frame,
which is valid in the non-relativisitic region $\beta\ll 1$.
Until now, no analytic evaluation of the full relativistic scattering amplitude at the two-loop level has been available.
Hence, we aim to devise a method to systematically evaluate this series expansion.
We also calculate the same scattering amplitude in pNRQCD EFT.
Then we determine the
color-singlet Hamiltonian in the latter theory by matching the two amplitudes
(``direct matching'').

Let us first highlight the characteristic features of our calculational method compared to the previous one. 
The calculation of the N$^3$LO Hamiltonian \cite{Kniehl:2002br} was, so to speak, done ``manually," utilizing the expansion-by-regions (EBR) technique \cite{Beneke:1997zp}. 
This technique was used to separate contributions from the hard (H) and soft (S) regions from those of the potential (P) and ultra-soft (US) regions, where 
only the H and S contributions constitute the Hamiltonian of the EFT.
This is not sufficient to determine the Hamiltonian unless we perform the matching off-shell.
In the EFT, there are different choices for the operator
basis of the Hamiltonian that are equivalent on-shell,
and depending on which one is chosen, the off-shell effects within the loop change.
Therefore, it is necessary to identify and manually add the operators (``off-shell operators'' \cite{Kniehl:2002br}) that compensate for these effects
and reproduce correctly the $S$-matrix elements.

However, it is highly demanded to systematize the calculational procedure to handle the vast amount of calculations required at higher orders. 
In particular, identifying  the off-shell operators
in the above manual procedure seems difficult to systematize at higher orders. 
In contrast, we systematize the calculation of the $\beta$ expansion with the aid of the differential equation satisfied by the master integrals, which enables calculation of contributions from the whole regions manageable.

Our procedure does not separate contributions from the S and P regions, although contributions from the P region eventually cancel out in the calculation of the Hamiltonian. 
This approach avoids the need to identify the off-shell operators,
at the cost of calculating
the P contributions which eventually cancel. 
(The direct matching procedure incorporates the off-shell operators and the cancellation of the P contributions simultaneously.)
In short, our method is adapted to systematically handle the large amount of calculations.

Moreover, we emphasize that this approach introduces a conceptually new methodology, enabling direct matching in a single step, as opposed to the conventional two-step process of integrating out each scale sequentially (QCD $\to$ NRQCD $\to$ pNRQCD).
As far as we understand, although the first step toward N$^4$LO calculations in the conventional approach (QCD $\to$ NRQCD) was completed some time ago \cite{Gerlach:2019kfo}, little advancement has been made since then.\footnote{
Partial results for the logarithmic part of the N$^4$LO Hamiltonian have been calculated, e.g.\ in
\cite{Penin:2004ay,Anzai:2018eua}.
} 
Our method breaks this impasse and paves the way for further developments.


We calculate the QCD two-loop on-shell scattering amplitude for
$Q(\vec{p})+\bar{Q}'(-\vec{p})\to Q(\vec{p}^{\,\prime})+\bar{Q}'(-\vec{p}^{\,\prime})$ in the 
color singlet channel and in the c.m.\ frame as follows,
where 
\bea
\vec{p}^{\,\prime}=\vec{p}+\vec{k}  \,,
~~~
\left| \vec{p} \right|^2 
=\left| \vec{p}^{\,\prime}\right|^2 
\,.
\eea
We first project the amplitude on to a spinor basis.
Feynman integrals are regularized by dimensional regularization, where the number of the space-time dimensions is set to $4-2\epsilon$. 
The spinor basis can be easily expressed by a two-component
spinor basis using the Pauli matrices in dimensional regularization,
satisfying
\bea
\{ \sigma^i, \sigma^j \}=2\delta^{ij}\mathbb{I}\,,
~~\delta^{ii}=3-2\epsilon\,,
~~{\rm tr}\,\mathbb{I}=2\,.
\eea
We adopt a 21-dimensional basis $\{\Lambda_1,\dots,\Lambda_{21}\}$, where
the first five elements are chosen as
\bea
&&
\Lambda_1=\mathbb{I}\otimes\mathbb{I}\,,
 \nonumber\\ &&
\Lambda_2=\sigma^a\,\sigma^b \otimes \sigma^a\,\sigma^b\,,
 \nonumber \\ &&
\Lambda_3=\sigma^a\,\sigma^b\, \sigma^c\,\sigma^d \otimes \sigma^a\,\sigma^b\, \sigma^c\,\sigma^d
\,,
 \\ &&
\Lambda_4=\frac{1}{m^2}\left(\vec{\sigma}\!\cdot\!\vec{k}\,\sigma^a \otimes \vec{\sigma}\!\cdot\!\vec{k}\,\sigma^a  \right)
 \,,
 \nonumber \\ &&
\Lambda_5=
\frac{1}{m^2}\left(\vec{\sigma}\!\cdot\!\vec{p}^{\,\prime}\,\vec{\sigma}\!\cdot\!\vec{p} \otimes \mathbb{I}
+ \mathbb{I}\otimes \vec{\sigma}\!\cdot\!\vec{p}^{\,\prime}\,\vec{\sigma}\!\cdot\!\vec{p}\right)
\,.
\nonumber
\eea
$m$ denotes the pole mass of the heavy quarks,
and the amplitude is renormalized in the on-shell scheme.
The coefficients of the spinor basis are expressed by scalar integrals by the projection.

Next we express the scalar integrals by master integrals using the integration-by-parts identities \cite{Chetyrkin:1981qh}.
We use the program \texttt{Kira} \cite{Maierhofer:2017gsa,Klappert:2020nbg,Fermat} 
(also \texttt{LiteRed} \cite{Lee:2013mka} and \texttt{FIRE} \cite{Smirnov:2014hma} for cross checks)
for this reduction, by which all the coefficients are expressed by 149
master integrals.
Up to this stage the obtained expression is exact.

We expand the master integrals in $\beta$.
This is done by solving the differential equation
satisfied by the master integrals \cite{Gehrmann:1999as}
in series expansions in $\beta$, that is, expansions
in $p$ and $k$.
We can interpret the $\beta$ expansions
in the language of the EBR technique, where
non-zero contributions originate from seven regions:
HH, HS, HP, SS, SP, PP, and PUS regions.
We can group them into four regions as
HH, HS+HP, SS+SP+PP and PUS.
Then the differential equation is satisfied independently by
each of the contributions from these four regions.\footnote{
The contributions from the S and P regions mix with each other
through the prescription in the EBR technique \cite{Beneke:1997zp},
which subtracts the pinch singularities in the S region
and compensates them as regularized contributions
in the P region.
As a result, the contributions from HS, HP, SS, SP and PP regions individually do not satisfy the differential equation.
}
We determine the boundary condition for the solution corresponding to each
region by
evaluating the LO term of the expansion using the EBR technique.
Moreover, in many cases, the boundary condition can be fixed by simply
demanding regularity of the solution at 
$u=\vec{k}^2-4\vec{p}^{\,2}=0$ on the physical sheet of the complex plane.
Once we have the $\beta$ expansions of the master integrals, 
it is straightforward to expand the scattering amplitude in $\beta$.
This procedure is much more efficient and economical
than to calculate each master integral
or each diagram by only the EBR technique.

The structure of the QCD $Q\bar{Q}'$ scattering amplitude is fairly complicated,
even after the expansion in $\beta$.
Although the amplitude is
regular at $u=0$ on the physical sheet, it has singularities at $u=0$
on the second and other Riemann sheets by analytical continuation.
[An example of such a structure is $\log(4\vec{p}^{\,2}/\vec{k}^2)/(\vec{k}^2-4\vec{p}^{\,2})$.]
The regularity of the amplitude
at $u=0$ on the physical sheet follows from the flavor conservation
of QCD.
Furthermore, the contributions from the P region include non-elementary functions 
of $p/k$ (counted as order one in the $\beta$ expansion).

The calculational procedure for the scattering amplitude for the same
process in pNRQCD EFT is
similar.\footnote{
We have not calculated the contributions from the US regions on the EFT side and simply assumed that they are the
same as the PUS contribution of QCD.
This is expected since in QCD the PUS contribution does not mix with other contributions and 
therefore is well
defined.
}
The amplitude is projected on to the two-component spinor basis.
The scalar integrals are reduced to master integrals of the EFT using the integration-by-parts
identities.
After matching the amplitude to that of QCD, we readily obtain the two-loop
Hamiltonian in the $\beta$ expansion.
The Hamiltonian up to ${\cal O}(\beta^0)$
consists only of the five operators $\Lambda_1,\dots,\Lambda_5$
of the spinor basis before expansion in $\epsilon$, where the coefficients of the other operators vanish.

We have performed the following cross checks for our results
before expanding them in $\epsilon$.
(1) We checked that the QCD scattering amplitude is
regular at $u=0$ on the physical sheet, even though this is
not so obvious from the obtained expression, which has singularities at $u=0$
on the second and other Riemann sheets and contains non-elementary functions
of $p/k$.
(2) 
The two-loop Hamiltonian is regular in $\vec{p}$ but singular in $\vec{k}$. 
Each term of the coefficient of $\Lambda_i$
takes the form $P_{1}(\vec{p})V(k)P_{2}(\vec{p}^{\,\prime})$, where $P_{j}$ is a homogeneous polynomial and $V(k)$ is proportional to $k^{a+b\epsilon}$ with $a\in \{-2,-1,0\}$, $b\in \{-4,-2,0\}$.
This form is expected to originate from the hard (H) and soft (S) contributions according to the EBR technique and is consistent with the concept of the EFT construction.
This means that all the singularities at $u=0$ on the second and other Riemann sheets 
cancel between the scattering amplitudes of QCD and pNRQCD EFT. 
Additionally, the functions of $p/k$ (P contributions) also cancel out.

After expanding our results in $\epsilon$, we have performed
the following cross checks.
(3)
We reproduced the known two-loop static QCD potential
$V_{\rm QCD}^{\text{(2 loop)}}$ \cite{Schroder:1999sg} and two-loop $1/(mk)$ potential
$V_{1/(mk)}^{\text{(2 loop)}}$ \cite{Kniehl:2001ju}
as part of the two-loop Hamiltonian. 
We also reproduced
the known one-loop part of the N$^3$LO Hamiltonian \cite{Kniehl:2002br}
at an intermediate stage of the calculation.
(4) We evaluated the coefficients of the expansion in $\epsilon$ of
each QCD master integral by numerical integrations 
(in part using \texttt{FIESTA} \cite{Smirnov:2021rhf,Borinsky:2020rqs})
in the non-relativistic region.
We compared them
with the expansion of the master integral in $\beta$ and $\epsilon$ 
obtained analytically
and checked consistency.
(5) Some of the $\epsilon$ expansions of
the master integrals are known analytically \cite{Smirnov:2001cm,Bonciani:2003cj}.
We expanded the analytical expressions in $\beta$ and found agreement with our results.

Let us present our final result.
We expand the coefficients of $\Lambda_1,\dots,\Lambda_5$ in $\epsilon$
while we ignore any ${\cal O}(\epsilon)$ contributions in $\Lambda_i$.
(Namely, we simply take the limit $\epsilon\to 0$ for $\Lambda_i$.)
The Hamiltonian is given in the form
\bea
&&
H=\frac{16\pi^2 C_F }{ k^2} \sum_{i=1}^4 \sum_{j=1}^3
\sum_{n,\ell\ge 0}\left(\frac{\alpha_s(k)}{4\pi}\right)^j 
 \nonumber\\ &&
~~~~~~~~~
\times
C_{\{i,j,n,2\ell\}} \left(\frac{k}{m}\right)^n 
\frac{(p^2)^\ell+(p'^2)^\ell}{2\,m^{2\ell}}
\, O_i\,,
\eea
where
\bea
p^2=
\left| \vec{p} \right|^2 
\,,~~
p'^2=\left| \vec{p}^{\,\prime}\right|^2 \,,
~~
k=\bigl| \vec{k} \bigr|=
\left| \vec{p}^{\,\prime}-\vec{p} \right|
\,.
\eea
We absorb $\log(\mu/k)$ terms originating from
the running of $\alpha_s$ by expressing the Hamiltonian by
$\alpha_s(k)$, the strong coupling constant in the 
modified minimal subtraction ($\overline{\rm MS}$) scheme of the theory with
$n_l$ flavors only, renormalized at $\mu=k$.\footnote{
In expressing
the Hamiltonian, it is customary to rewrite 
the coupling constant of the full theory $\alpha_s^{(n_h+n_l)}(\mu)$ by that of
the theory with
$n_l$ flavors only $\alpha_s(\mu)\equiv \alpha_s^{(n_l)}(\mu)$.
We include the ${\cal O}(\epsilon)$ term in this decoupling relation \cite{Grozin:2007fh,Grozin:2011nk},
which simplifies the result slightly since the one-loop Hamiltonian includes the $1/\epsilon$ pole.
}\footnote{
We do not include the ${\cal O}(\epsilon)$ correction
to the running formula
$$
\alpha_s(k)=\alpha _s(\mu )-
\frac{\alpha _s(\mu
   ){}^2}{4 \pi} \left(11-\frac{ 2 n_l}{3}\right) \log \left(\frac{k^2}{\mu ^2}\right)+\cdots
   \,.
$$
}
The spinor basis in three dimensions is defined as
\bea
&&
O_1=\mathbb{I}\otimes \mathbb{I}\,,
~
O_2=\vec{S}^2 \,,
~
O_3=\frac{i}{k^2}\,\vec{S}\cdot\left(\vec{p}\times\vec{k}\right)\,,
 \nonumber\\ &&
~~
O_4=\sigma^a \otimes \sigma^a-\frac{3}{k^2}\left(\vec{k}\cdot\vec{\sigma}\right)\otimes\left(\vec{k}\cdot\vec{\sigma}\right)\,,
\eea
with 
\bea
\vec{S}=\frac{\vec{\sigma}}{2}\otimes \mathbb{I} + \mathbb{I}\otimes\frac{\vec{\sigma}}{2} \,.
\eea
The Wilson coefficients are separated into finite and divergent parts as
\bea
C_{\{i,j,n,2\ell\}}=C^{\text{fin}}_{\{i,j,n,2\ell\}}+C^{\text{div}}_{\{i,j,n,2\ell\}} \,.
\eea
Two-loop non-zero finite Wilson coefficients up to ${\cal O}(\beta^0)$ are given by
\bea
&&
C^{\text{fin}}_{\{1,3,0,0\}}= V_{\rm QCD}^{\text{(2 loop)}}(k) \cdot ({4\pi k^2})/ [{C_F \alpha_s(k)^3} ] ,
 \\ &&
C^{\text{fin}}_{\{1,3,1,0\}}=V_{1/(mk)}^{\text{(2 loop)}}(k) \cdot ({4\pi m k})/[{C_F \alpha_s(k)^3} ] ,
\eea
\bea
&&
C^{\text{fin}}_{\{1,3,2,0\}}=
L_{m/\mu}
\left(-\frac{208}{9}-\frac{260 \pi ^2}{9}+\frac{40 n_l}{27}\right) 
 \nonumber\\ && ~~~~~
+
L_{k/m}
\biggl( -\frac{106 \pi ^2}{9}
-\frac{2152}{9}
+\frac{3013 n_l}{162}
-\frac{17 n_h}{6}\biggr)
 \nonumber\\ && ~~~~~
+L_{k/m}^2
\biggl(-\frac{1825}{18}+\frac{109 n_l}{27} \biggr)
+\frac{27 \pi ^4}{16}
+\frac{1285 \zeta
   (3)}{6}
 \nonumber\\ && ~~~~~
   -\frac{11}{3}\pi^2\log (2) 
+\frac{99541 \pi ^2}{1944}
+\frac{41395}{162} 
+\frac{8}{27} n_h n_l
 \nonumber\\ && ~~~~~
+n_h \left(\frac{277 \pi
   ^2}{324}-\frac{1375}{108}\right)
+n_l\left(\frac{139}{81}+\frac{220 \pi ^2}{81}\right)
\,,
 \nonumber\\
\\
&&
C^{\text{fin}}_{\{1,3,0,2\}}=
L_{k/\mu}
\left(32 \pi ^2+416-\frac{80  n_l}{3}\right)
-\frac{100 n_l^2}{81}
 \nonumber\\ && ~~~~~
+n_l \biggl(\frac{52 \zeta
   (3)}{3}
   +\frac{1901}{27}-\frac{4 \pi ^2}{9}\biggr)
-\frac{9 \pi ^4}{4}
-114\zeta(3)
 \nonumber\\ && ~~~~~
+\frac{266 \pi ^2}{3}-\frac{7919}{18}
\,,
\\
&&
C^{\text{fin}}_{\{2,3,2,0\}}=
\frac{352}{27} \pi ^2
L_{m/\mu}+
L_{k/m}^2
\left(38-\frac{5 n_l}{3}\right)
 \nonumber\\ && ~~~~~
+
L_{k/m}\biggl(\frac{584}{9}-\frac{182 n_l}{27}
-4 \pi ^2\biggr)
+
\frac{100 n_l^2}{243}
 \nonumber\\ && ~~~~~
+n_l \biggl(-\frac{52 \zeta
   (3)}{9}
   -\frac{1777}{81}-\frac{10 \pi ^2}{9}\biggr)
-\frac{3 \pi
   ^4}{4}  
 \nonumber\\ && ~~~~~
    -\frac{1181 \zeta (3)}{6}
   -\frac{121}{9} \pi ^2\log (2)
   -\frac{1133 \pi ^2}{81}
+\frac{10771}{54}
 \nonumber\\ && ~~~~~
+   n_h
\left(\frac{770}{81}-\frac{16 \pi ^2}{27}\right)
\,,
\\
&&
C^{\text{fin}}_{\{3,3,2,0\}}=
L_{k/m}^2 (42-2n_l) +L_{k/m}
\biggl(\frac{1010}{3}
-\frac{242 n_l}{9}\biggr) 
 \nonumber\\ && ~~~~~
+n_l \left(-26 \zeta (3)
-\frac{4591}{54}-\frac{4 \pi
   ^2}{3}\right)
-\frac{27 \pi ^4}{8}
 \nonumber\\ && ~~~~~
+\frac{161 \zeta (3)}{3}
+\frac{56}{9} \pi ^2 \log (2)
+\frac{820 \pi
   ^2}{27}
   +\frac{7823}{12}
 \nonumber\\ && ~~~~~
   +\frac{50n_l^2}{27}
+n_h\biggl(\frac{1010}{27}
-\frac{34 \pi ^2}{9}\biggr) 
\,,
\\
&&
C^{\text{fin}}_{\{4,3,2,0\}}=
L_{k/m}^2 \biggl(\frac{17}{4}-\frac{n_l}{6}\biggr)
+L_{k/m}\biggl(\frac{583}{18}
-\frac{121 n_l}{54}\biggr)
 \nonumber\\ && ~~~~~
+n_l\left(-\frac{13 \zeta (3)}{9}-\frac{2257}{324}-\frac{\pi
   ^2}{9}\right)
-\frac{3 \pi ^4}{16}+\frac{31 \zeta (3)}{18}
 \nonumber\\ && ~~~~~
+\frac{14}{27} \pi ^2 \log (2)
+\frac{545 \pi
   ^2}{162}
+\frac{15739}{216}
   +\frac{25
   n_l^2}{243}
 \nonumber\\ && ~~~~~
   +n_h \biggl(\frac{505}{162}
-\frac{17 \pi ^2}{54}\biggr) 
\,,
\eea
where $L_{a/b}$ represents $\log(a^2/b^2)$.
Note that
$V_{\rm QCD}^{\text{(2 loop)}}$ and
$V_{1/(mk)}^{\text{(2 loop)}}$ do not include the heavy-quark-loop contributions, which
are included in other Wilson coefficients.

Two-loop non-zero divergent Wilson coefficients are given by
\bea
&&
C_{\{1,3,1,0\}}^{\text{div}}=
-\frac{136 \pi ^2}{3 \epsilon }
\,,
\\ && 
C_{\{1,3,2,0\}}^{\text{div}}=
L_{k/\mu}
\left(\frac{44}{9 \epsilon }-\frac{8 n_l}{27 \epsilon }\right)
-\frac{22}{9 \epsilon ^2}
 \nonumber\\ && ~~~~~
+\frac{130 \pi ^2}{9 \epsilon }+\frac{104}{9
   \epsilon }
   +n_l \left(\frac{4}{27 \epsilon ^2}-\frac{20}{27
   \epsilon }\right)
\,,
\\ 
&& 
C_{\{1,3,0,2\}}^{\text{div}}=
L_{k/\mu}
\left(-\frac{88}{\epsilon }+\frac{16 n_l}{3 \epsilon }\right)
+\frac{44}{\epsilon ^2}
 \nonumber\\ && ~~~~~
-\frac{16 \pi ^2}{\epsilon }-\frac{208}{\epsilon }
+n_l\left(\frac{40}{3 \epsilon }-\frac{8}{3 \epsilon
   ^2}\right)
\,,
\\ 
&& 
C_{\{2,3,2,0\}}^{\text{div}}=
-\frac{176 \pi ^2}{27 \epsilon }
\,.
\eea

In summary we computed the quarkonium Hamiltonian at the two-loop level
in the non-annihilation channel.
The obtained Hamiltonian has an expected form 
as resulting from integrating the H and S modes.
This shows that the
singular structure of the scattering amplitude at $u=0$ on the
second and other Riemann sheets as well as non-elementary functions
of $p/k$ originating from the P region are reproduced
correctly by the EFT
and canceled in the calculation of the Hamiltonian.
The developed calculational procedure would also be useful to 
compute various Wilson coefficients at high orders
relevant for quarkonium observables,
including a straightforward application to the calculation of the annihilation channel.
In particular, the obtained Hamiltonian will play a major role, for instance, in calculating
the fine and hyperfine splittings of the quarkonium.

The structure of the scattering amplitude near the threshold of fermion pairs is highly complex, yet the fact that it can be described by the simple form of the Hamiltonian obtained in this paper, and the pNRQCD effective theory utilizing it, can be considered to be highly non-trivial. This effective theory allows for a clear understanding of the analytic structure in terms of Green functions of quantum mechanics.
Thus, our result is expected to contribute to future analyses of the structure of amplitudes near the threshold in processes such as Bhabha scattering and quark-antiquark scattering, where such analyses are already difficult at two-loop level up to now. (For the current status of the study of the full two-loop scattering amplitude in the case of QED, see Ref.~\cite{Delto:2023kqv}.)

For a consistent calculation of physical observables at N$^4$LO accuracy
using the Hamiltonian, it is
sometimes required to include higher-order terms of $\beta$ and
$\epsilon$ at the tree and one-loop levels than those given in the literature.
We provide them as well as the expression of the two-loop Hamiltonian
for general color factors in the Supplementary Material \cite{SupplementaryMat}.

\section*{Acknowledgement}
The works of Y.S.\ and H.T., respectively, were supported in part
by JSPS KAKENHI Grant Numbers JP23K03404 and
by JP19K14711 and JP23K13110.
H.T.\ is the Yukawa Research Fellow supported by Yukawa Memorial Foundation.




%
%
%
%
%
%
%

\newpage
\begin{widetext}
\oddsidemargin0.3cm
\topmargin-0.5cm
\textheight22.5cm
\textwidth16cm
~
\begin{center}
\Large\bf
Supplemental: More details of Hamiltonian 
\end{center}

\large
We present the tree-level and one-loop Hamiltonians
before expanding in $\epsilon$,
up to ${\cal O}(\beta^2)$ and ${\cal O}(\beta)$
[${\cal O}(\beta^4)$ and ${\cal O}(\alpha_s \beta^3)$ relative to the LO $\sim \alpha_s/\beta^2$], 
respectively.
These are in general necessary ingredients to calculate physical observables at the N$^4$LO
accuracy.
We also present the two-loop Hamiltonian before and after the expansion in $\epsilon$ and retaining
the color factors. (The former is given as an electronic file.)

\appendix
\section{\large Tree-level and one-loop Hamiltonians}

The Hamiltonian is given in the form
\bea
H=\frac{C_F \,\bar{\mu}^{-2\epsilon}}{ k^2} \sum_{i=1}^6 \sum_{j=1}^3
\sum_{n,\ell\ge 0}\left(g_R^2 \,\bar{\mu}^{2\epsilon}\right)^j W_{\{i,j,n,2\ell\}} \left(\frac{k}{m}\right)^n 
\frac{(p^2)^\ell+(p'^2)^\ell}{2\,m^{2\ell}}\, \Lambda_i\,,
\label{defW}
\eea
where 
\bea
k=\left| \vec{k} \right| \,,
~~~
p^2=\left| \vec{p} \right|^2 \,,
~~~
p'^2=\left| \vec{p}^{\,\prime}\right|^2 \,,
~~~
\vec{p}^{\,\prime}=\vec{p}+\vec{k}  \,.
\eea
$g_R=\sqrt{4\pi \alpha_s^{(n_h+n_l)}(\mu)}$ denotes the renormalized gauge coupling constant in the $\overline{\rm MS}$ scheme of the full theory (with $n_h$ heavy quark flavors and
$n_l$ massless quark flavors);
$\bar{\mu}^2=\mu^2 \,e^{\gamma_E}/(4\pi) $, where $\gamma_E=0.5772\dots$ denotes the Euler constant.
The spinor basis is defined in dimensional regularization as
\bea
&&
\Lambda_1=\mathbb{I}\otimes\mathbb{I}\,,
 \nonumber\\ &&
\Lambda_2=\sigma^a\,\sigma^b \otimes \sigma^a\,\sigma^b\,,
 \nonumber \\ &&
\Lambda_3=\sigma^a\,\sigma^b\, \sigma^c\,\sigma^d \otimes \sigma^a\,\sigma^b\, \sigma^c\,\sigma^d
\,,
 \\ &&
\Lambda_4=\frac{1}{m^2}\left(\vec{\sigma}\!\cdot\!\vec{k}\,\sigma^a \otimes \vec{\sigma}\!\cdot\!\vec{k}\,\sigma^a  \right)
 \,,
 \nonumber \\ &&
\Lambda_5=
\frac{1}{m^2}\left(\vec{\sigma}\!\cdot\!\vec{p}^{\,\prime}\,\vec{\sigma}\!\cdot\!\vec{p} \otimes \mathbb{I}
+ \mathbb{I}\otimes \vec{\sigma}\!\cdot\!\vec{p}^{\,\prime}\,\vec{\sigma}\!\cdot\!\vec{p}\right)
\,,
\nonumber \\ &&
\Lambda_6=\frac{1}{m^4}\left(\vec{\sigma}\!\cdot\!\vec{p}^{\,\prime}\,\vec{\sigma}\!\cdot\!\vec{p} \otimes  \vec{\sigma}\!\cdot\!\vec{p}^{\,\prime}\,\vec{\sigma}\!\cdot\!\vec{p}\right) 
\,.
\nonumber
\eea
We present the Wilson coefficients before expansion in $\epsilon$.
We list only those coefficients which are non-zero.

At tree level and  up to ${\cal O}(\beta^4)$ relative to LO, they are given by
\bea
&&
W_{\{1,1,0,0\}}= -1,
~~~
W_{\{1,1,0,2\}}=\frac{1}{2}\, ,
~~~
W_{\{1,1,0,4\}}=-\frac{7}{16}\, ,
\nonumber \\ &&
W_{\{4,1,0,0\}}= -\frac{1}{4},
~~~
W_{\{4,1,0,2\}}=\frac{1}{4}\, ,
~~~
W_{\{5,1,0,0\}}=-\frac{3}{4}\, ,
~~~
W_{\{5,1,0,2\}}=\frac{3}{4}\, ,
\\ &&
W_{\{6,1,0,0\}}= -\frac{1}{16} \,.
\nonumber 
\eea
At one loop  and  up to ${\cal O}(\alpha_s\beta^3)$ relative to LO, the Wilson coefficients are given by
\bea
&&
W_{\{1,2,0,0\}}=
m^{-2 \epsilon } \cdot \frac{2}{3}  (\epsilon -1) n_h\, iI_H
-\bar{\mu}^{-2\epsilon}\cdot 2\, \delta _1 Z_g 
\nonumber \\ && ~~~~~~~~     
+k^{-2 \epsilon } \left(-\frac{(\epsilon -1) (8 \epsilon -11) C_A}{2
   \epsilon -3}-\frac{2 (\epsilon -1) n_l}{2 \epsilon
   -3}\right) iI_S^a
\,,
\\&&
W_{\{1,2,1,0\}}=
k^{-2 \epsilon }  \left((\epsilon -1) C_A-\frac{1}{2} (2
   \epsilon -1) C_F\right)iI_S^b
\,,
 \\ && 
W_{\{1,2,2,0\}}=
m^{-2\epsilon} \biggl(-\frac{(\epsilon -1) \left(96 \epsilon ^3-100 \epsilon
   ^2+12 \epsilon -29\right) C_A}{24 (2 \epsilon -1) (2 \epsilon
   +1)}
\nonumber \\ && ~~~~~~~~     
   +\frac{(\epsilon -1) \left(96 \epsilon ^4+44 \epsilon ^3-96 \epsilon
   ^2+37 \epsilon +6\right) C_F}{6 (2 \epsilon -1) (2 \epsilon +1) (2
   \epsilon +3)}
   -\frac{2}{15} \epsilon(\epsilon -1)   n_h\biggr)iI_H
\nonumber \\ && ~~~~~~~~  
   +k^{-2
   \epsilon } \left(\frac{1}{24} \left(-48 \epsilon ^2+104
   \epsilon -61\right) C_A+\frac{1}{3} (\epsilon -1) (8 \epsilon -7)
   C_F\right) iI_S^a
\,,
 \\ && 
W_{\{1,2,0,2\}}=
m^{-2\epsilon} 
 \left(\frac{2 (\epsilon -1) \left(2 \epsilon ^2-1\right)
   C_A}{2 \epsilon -1}-\frac{4 (\epsilon -1) \epsilon  (2 \epsilon +1)
   C_F}{2 \epsilon -1}+\frac{1}{3} (1-\epsilon ) n_h\right)iI_H
\nonumber \\ && ~~~~~~~~  
   +\bar{\mu}^{-2\epsilon}\, \delta _1 Z_g
   +k^{-2 \epsilon }
\left(\frac{(\epsilon -1) n_l}{2 \epsilon
   -3}-\frac{\left(40 \epsilon ^2-95 \epsilon +51\right) C_A}{6 (2 \epsilon
   -3)}\right) iI_S^a
\,,
 \\ && 
W_{\{1,2,1,2\}}=
k^{-2 \epsilon } \left(\frac{1}{2} (\epsilon -2)
   C_A+\frac{1}{4} (-\epsilon -1) C_F\right) {iI}_S^b 
\,,
 \\ && 
W_{\{1,2,3,0\}}=
k^{-2 \epsilon } \left(\frac{1}{8} \left(2 \epsilon ^2-6
   \epsilon +5\right) C_A+\frac{1}{16} \left(-6 \epsilon ^2+15 \epsilon
   -14\right) C_F\right) {iI}_S^b
\,,
 \\ && 
W_{\{2,2,2,0\}}=
m^{-2\epsilon} 
    \left(\frac{1}{8} (1-\epsilon ) C_A+\frac{(\epsilon -1)
   \epsilon  C_F}{2 (2 \epsilon +1)}\right)i I_H
   -k^{-2 \epsilon }\cdot\frac{1}{8} C_A 
   i I_S^a
\,,
 \\ && 
W_{\{2,2,3,0\}}=
-k^{-2 \epsilon }\cdot\frac{\epsilon  C_F }{16 (\epsilon -1)}\, i I_S^b
\,,
 \\ && 
W_{\{4,2,0,0\}}=
m^{-2\epsilon} 
 \left(-\frac{(\epsilon -1) \left(2 \epsilon ^2-1\right) C_A}{2
   (2 \epsilon -1)}+\frac{(\epsilon -1) \epsilon  (2 \epsilon +1) C_F}{2
   \epsilon -1}+\frac{1}{6} (\epsilon -1) n_h\right)i I_H
\nonumber \\ && ~~~~~~~~  
   -\bar{\mu}^{-2\epsilon} \,\frac{\delta _1 Z_g}{2}
   +k^{-2 \epsilon }
    \left(-\frac{(\epsilon -1) (4 \epsilon -5) C_A}{4 (2
   \epsilon -3)}-\frac{(\epsilon -1) n_l}{2 (2 \epsilon
   -3)}\right)i I_S^a
\,,
 \\ && 
W_{\{4,2,1,0\}}=
k^{-2 \epsilon } \left(\frac{\epsilon  C_A}{8}-\frac{\left(2
   \epsilon ^2-7 \epsilon +4\right) C_F}{16 (\epsilon -1)}\right) i I_S^b 
\,,
\\ && 
W_{\{5,2,0,0\}}=
m^{-2\epsilon} 
\left(-\frac{(\epsilon -1) \left(2 \epsilon ^2-1\right) C_A}{2
   \epsilon -1}+\frac{2 (\epsilon -1) \epsilon  (2 \epsilon +1) C_F}{2
   \epsilon -1}+\frac{1}{2} (\epsilon -1) n_h\right)i I_H 
\nonumber \\ && ~~~~~~~~  
   -\bar{\mu}^{-2\epsilon} \,   \frac{3 \delta _1 Z_g}{2}
   +k^{-2 \epsilon }
   \left(-\frac{\left(24 \epsilon ^2-49 \epsilon +21\right)
   C_A}{4 (2 \epsilon -3)}-\frac{3 (\epsilon -1) n_l}{2 (2 \epsilon
   -3)}\right)i I_S^a 
\,,
 \\ && 
W_{\{5,2,1,0\}}=
   k^{-2 \epsilon } \left(\frac{1}{4} (3 \epsilon -2)
   C_A+\frac{1}{4} (4-3 \epsilon ) C_F\right)i I_S^b 
   \,.
\eea
The master integrals of the hard and soft regions in the expansion-by-regions
technique are given by
\bea
&&
i I_H=
(4 \pi )^{\epsilon -2} \Gamma (\epsilon -1)
\,,
~~~
iI_S^a=
-\frac{2^{4 \epsilon -5} \pi ^{\epsilon -\frac{1}{2}} }{\sin (\pi  \epsilon
   )
\Gamma \left(\frac{3}{2}-\epsilon \right)}
\,,
~~~
iI_S^b=
\frac{16^{\epsilon -1} \pi ^{\epsilon } }{\cos (\pi  \epsilon )\Gamma
   (1-\epsilon )}
\,, 
~~~
\eea
(after factoring out the dimensionful parameters).
The one-loop counter term for the gauge coupling constant reads
\bea
&& 
\delta _1 Z_g=
\frac{2 \left(n_h+n_l\right)-11 C_A}{96 \pi ^2 \epsilon }
\,. 
\eea
The color factors are given by $C_F=4/3$ and $C_A=3$ for the ${\rm SU}(3)$ gauge group.
It is straightforward to expand the above Wilson coefficients in $\epsilon$.

\section{\large Two-loop Hamiltonian for general gauge group}

We expand the coefficients\footnote{
$\Lambda_6$ is ${\cal O}(\beta^4)$ by itself, and it contributes only to the tree-level Hamiltonian
within the accuracy orders of our current interest.
} 
of $\Lambda_1,\dots,\Lambda_5$ in $\epsilon$
while we ignore any ${\cal O}(\epsilon)$ contributions in $\Lambda_i$.
(Namely, we simply take the limit $\epsilon\to 0$ for $\Lambda_i$.)
The Hamiltonian is given in the form
\bea
H=\frac{16\pi^2 C_F }{ k^2} \sum_{i=1}^4 \sum_{j=1}^3
\sum_{n,\ell\ge 0}\left(\frac{\alpha_s(k)}{4\pi}\right)^j C_{\{i,j,n,2\ell\}} \left(\frac{k}{m}\right)^n 
\frac{(p^2)^\ell+(p'^2)^\ell}{2\,m^{2\ell}}
\, O_i\,.
\eea
$\alpha_s(k)$ denotes the strong coupling constant in the $\overline{\rm MS}$ scheme of the theory with
$n_l$ flavors only, renormalized at $\mu=k$.\footnote{
It is customary to express the Hamiltonian in terms of
the coupling constant of 
the theory with
$n_l$ flavors only $\alpha_s(\mu)\equiv \alpha_s^{(n_l)}(\mu)$.
We include the ${\cal O}(\epsilon)$ term in the decoupling relation \cite{Grozin:2007fh,Grozin:2011nk}
to rewrite $\alpha_s^{(n_h+n_l)}(\mu)$ by $\alpha_s(\mu)$.
}$^,$\footnote{
We do not include the ${\cal O}(\epsilon)$ correction
to the running formula
$$
\alpha_s(k)=\alpha _s(\mu )-
\frac{\alpha _s(\mu
   ){}^2}{4 \pi} \left(11-\frac{ 2 n_l}{3}\right) \log \left(\frac{k^2}{\mu ^2}\right)+\cdots
   \,.
$$
}
The spinor basis in three dimensions is defined as
\bea
&&
O_1=\mathbb{I}\otimes\mathbb{I}\,,
~~
O_2=\vec{S}^2 \,,
~~
O_3=\frac{i}{k^2}\,\vec{S}\cdot\left(\vec{p}\times\vec{k}\right)\,,
~~
O_4=\sigma^a \otimes \sigma^a-\frac{3}{k^2}\left(\vec{k}\cdot\vec{\sigma}\right)\otimes\left(\vec{k}\cdot\vec{\sigma}\right)\,,
~~~
\nonumber\\
\eea
with 
\bea
\vec{S}=\frac{\vec{\sigma}}{2}\otimes \mathbb{I} + \mathbb{I}\otimes\frac{\vec{\sigma}}{2} \,.
\eea
The Wilson coefficients are separated into finite and divergent parts as
\bea
C_{\{i,j,n,2\ell\}}=C^{\text{fin}}_{\{i,j,n,2\ell\}}+C^{\text{div}}_{\{i,j,n,2\ell\}}
\,.
\eea
2-loop non-zero finite Wilson coefficients are given by
\bea
&&
C^{\text{fin}}_{\{1,3,0,0\}}= V_{\rm QCD}^{\text{(2 loop)}}(k) \, [{C_F \alpha_s(k)^3}/({4\pi k^2}) ]^{-1}
\nonumber \\ && ~~~~~~~~ ~~~
=-\frac{100 n_l^2}{81}+n_l \left(\left(\frac{28 \zeta (3)}{3}+\frac{899}{81}\right) C_A+\left(\frac{55}{6}-8 \zeta
   (3)\right) C_F\right)
\nonumber \\ && ~~~~~~~~ ~~~~~
   +\left(-\frac{22 \zeta (3)}{3}-\frac{4343}{162}-4 \pi ^2+\frac{\pi
   ^4}{4}\right) C_A^2
\, ,
 \\ &&
C^{\text{fin}}_{\{1,3,1,0\}}=V_{1/(mk)}^{\text{(2 loop)}}(k) \, [{C_F \alpha_s(k)^3}/({4\pi m k}) ]^{-1}
\nonumber \\ && ~~~~~~~~ ~~~
=\left(-\frac{32}{3} \pi ^2 C_A C_F-\frac{16}{3} \pi ^2 C_A^2\right) \log \left(\frac{\mu
   ^2}{k^2}\right)+n_l \left(\frac{49 \pi ^2 C_A}{18}-\frac{4 \pi ^2
   C_F}{9}\right)
\nonumber \\ && ~~~~~~~~ ~~~~~
   +\left(\frac{130 \pi ^2}{9}-\frac{32}{3} \pi ^2 \log (2)\right) C_A
   C_F+\left(-\frac{101 \pi ^2}{9}-\frac{16}{3} \pi ^2 \log (2)\right) C_A^2
\, ,
\\
&&
C^{\text{fin}}_{\{1,3,2,0\}}=
\log ^2\left(\frac{k^2}{m^2}\right) \left(-\frac{11 C_A C_F}{9}+\frac{13 C_A n_l}{9}-\frac{193 
C_A^2}{18}-\frac{2 C_F n_l}{9}\right)
\nonumber \\ && ~~~~~~~~ 
+
\log \left(\frac{k^2}{m^2}\right) \Biggl( -2 \pi ^2 C_F^2
+\left(\frac{146}{9}+\frac{13 \pi ^2}{9}\right) C_A C_F-\frac{17 C_A 
n_h}{18}+\frac{637 C_A n_l}{54}
\nonumber \\ && ~~~~~~~~ 
~~~~~
+\left(-\frac{304}{9}-\frac{14 \pi 
^2}{9}\right) C_A^2-\frac{340 C_F n_l}{27}\Biggr)
\nonumber \\ && ~~~~~~~~ + 
\log \left(\frac{m^2}{\mu ^2}\right) \Biggl(\left(\frac{416}{9}-\frac{13 \pi ^2}{9}\right) C_A C_F+\frac{40 C_A n_l}{9}-\frac{80 C_F n_l}{9}-4 \pi ^2 C_F^2
\nonumber \\ && ~~~~~~~~ 
~~~~~
+\left(-\frac{208}{9}-\frac{16 \pi ^2}{9}\right) C_A^2\Biggr)
\nonumber \\ && ~~~~~~~~ + 
\log (2) \left(22 \pi ^2 C_A C_F-9 \pi ^2 C_A^2-6 \pi ^2 C_F^2\right)
+
\left(\frac{103 \zeta (3)}{2}-\frac{407}{6}+\frac{13 \pi ^2}{24}+\frac{3 \pi ^4}{16}\right) C_A^2
\nonumber \\ && ~~~~~~~~ +
\left(-77 \zeta (3)+\frac{2369}{9}-\frac{713 \pi ^2}{54}\right) C_A C_F
+
\left(33 \zeta (3)-\frac{946}{9}+\frac{6023 \pi ^2}{108}\right) C_F^2
+
\frac{8 n_h n_l}{27}
\nonumber \\ && ~~~~~~~~ +
\left(\frac{1387}{108}-\frac{41 \pi ^2}{36}\right) C_A n_h
+
\left(\frac{173 \pi ^2}{54}-\frac{346}{9}\right) C_F n_h
+
\left(\frac{28 \pi ^2}{27}-\frac{173}{27}\right) C_A n_l
\nonumber \\ && ~~~~~~~~
+
\left(\frac{424}{27}-\frac{8 \pi ^2}{27}\right) C_F n_l
\,,
\\
&&
C^{\text{fin}}_{\{1,3,0,2\}}=
\log \left(\frac{k^2}{\mu ^2}\right) \left(\left(\frac{416}{9}+\frac{32 \pi ^2}{9}\right) C_A^2-\frac{80 C_A n_l}{9}\right)
+
\left(\frac{28 \zeta (3)}{3}
+\frac{1571}{81}-\frac{4 \pi ^2}{27}\right) C_A n_l
\nonumber \\ && ~~~~~~~~ 
+
\left(\frac{55}{6}-8 \zeta (3)\right) C_F n_l
-\frac{100 n_l^2}{81}
+
\left(-\frac{38 \zeta (3)}{3}-\frac{7919}{162}+\frac{266 \pi ^2}{27}-\frac{\pi ^4}{4}\right) C_A^2
\,,
\eea
\bea
&&
C^{\text{fin}}_{\{2,3,2,0\}}=
\log ^2\left(\frac{k^2}{m^2}\right) \left(\frac{38 C_A^2}{9}-\frac{5 C_A n_l}{9}\right)
+
\log \left(\frac{m^2}{\mu ^2}\right) \left(\frac{8}{3} \pi ^2 C_A C_F+\frac{4}{3} \pi ^2 C_F^2\right) 
\nonumber \\ && ~~~~~~~~ 
+
\log \left(\frac{k^2}{m^2}\right) \left(-\frac{46 C_A C_F}{9}-\frac{82 C_A n_l}{27}+\left(\frac{256}{27}-\frac{4 \pi ^2}{9}\right) C_A^2 \right.
+\frac{16 C_F n_l}{9}\Biggr)
\nonumber \\ && ~~~~~~~~ 
+
\log (2) \left(12 \pi ^2 C_F^2-\frac{238}{9} \pi ^2 C_A C_F+\frac{71}{9} \pi ^2 C_A^2\right)
+
\left(\frac{16 \pi ^2}{27}-\frac{538}{81}\right) C_A n_h
\nonumber \\ && ~~~~~~~~ 
+
\left(\frac{596}{27}-\frac{16 \pi ^2}{9}\right) C_F n_h
+
\frac{100 n_l^2}{243}
+
\left(-18 \zeta (3)-\frac{10}{3}-\frac{239 \pi ^2}{9}\right) C_F^2
\nonumber \\ && ~~~~~~~~ 
+
\left(-\frac{295 \zeta (3)}{18}+\frac{11651}{486}-\frac{127 \pi ^2}{27}-\frac{\pi ^4}{12}\right) C_A^2
+
\left(-\frac{13 \zeta (3)}{3}-\frac{70}{27}+\frac{170 \pi ^2}{9}\right) C_A C_F
\nonumber \\ && ~~~~~~~~ 
+
\left(-\frac{28 \zeta (3)}{9}-\frac{503}{243}-\frac{10 \pi ^2}{27}\right) C_A n_l
+
\left(\frac{8 \zeta (3)}{3}-\frac{637}{54}\right) C_F n_l
\,,
\\&&
C^{\text{fin}}_{\{3,3,2,0\}}=
\log \left(\frac{k^2}{m^2}\right) \left(\frac{56 C_A C_F}{3}-\frac{70 C_A n_l}{9}+\frac{262 C_A^2}{9}-\frac{8 C_F n_l}{3}\right)
\nonumber \\ && ~~~~~~~~ 
+
\log ^2\left(\frac{k^2}{m^2}\right) \left(\frac{14 C_A^2}{3}-\frac{2 C_A n_l}{3}\right)
+
\log (2) \left(\frac{8}{3} \pi ^2 C_A C_F+\frac{8}{3} \pi ^2 C_A^2-16 \pi ^2 C_F^2\right)
\nonumber \\ && ~~~~~~~~ 
+
\left(3 \zeta (3)+\frac{5999}{108}-\frac{4 \pi ^2}{9}-\frac{3 \pi ^4}{8}\right) C_A^2
+
\left(\frac{10 \pi ^2}{9}-\frac{298}{27}\right) C_A n_h
+
\left(\frac{476}{9}-\frac{16 \pi ^2}{3}\right) C_F n_h
\nonumber \\ && ~~~~~~~~ 
+
\frac{50 n_l^2}{27}
+
\left(-14 \zeta (3)-\frac{827}{54}-\frac{4 \pi ^2}{9}\right) C_A n_l
+
\left(-4 \zeta (3)+\frac{590}{9}+\frac{8 \pi ^2}{3}\right) C_A C_F
\nonumber \\ && ~~~~~~~~ 
+
\left(12 \zeta (3)-\frac{1055}{36}\right) C_F n_l
+
\left(24 \zeta (3)-62+\frac{40 \pi ^2}{3}\right) C_F^2
\,,
\\
&&
C^{\text{fin}}_{\{4,3,2,0\}}=
\log \left(\frac{k^2}{m^2}\right) \left(\frac{17 C_A C_F}{9}-\frac{35 C_A n_l}{54}+\frac{149 C_A^2}{54}-\frac{2 C_F n_l}{9}\right)
\nonumber \\ && ~~~~~~~~ 
+
\log ^2\left(\frac{k^2}{m^2}\right) \left(\frac{17 C_A^2}{36}-\frac{C_A n_l}{18}\right)
+
\log (2) \left(\frac{2}{9} \pi ^2 C_A C_F+\frac{2}{9} \pi ^2 C_A^2-\frac{4}{3} \pi ^2 C_F^2\right)
\nonumber \\ && ~~~~~~~~ 
+
\left(-\frac{\zeta (3)}{18}+\frac{12299}{1944}+\frac{\pi ^2}{18}-\frac{\pi ^4}{48}\right) C_A^2
+
\left(\frac{5 \pi ^2}{54}-\frac{149}{162}\right) C_A n_h
+
\left(\frac{119}{27}-\frac{4 \pi ^2}{9}\right) C_F n_h
\nonumber \\ && ~~~~~~~~ 
+
\frac{25 n_l^2}{243}
+
\left(-\frac{7 \zeta (3)}{9}-\frac{1367}{972}-\frac{\pi ^2}{27}\right) C_A n_l
+
\left(-\frac{\zeta (3)}{3}+\frac{331}{54}+\frac{2 \pi ^2}{9}\right) C_A C_F
\nonumber \\ && ~~~~~~~~ 
+
\left(\frac{2 \zeta (3)}{3}-\frac{445}{216}\right) C_F n_l
+
\left(2 \zeta (3)-\frac{29}{6}+\frac{10 \pi ^2}{9}\right) C_F^2
\,.
\eea
2-loop non-zero divergent Wilson coefficients are given by
\bea
&&
C_{\{1,3,1,0\}}^{\text{div}}=
-\frac{8 \pi ^2 C_A^2}{3 \epsilon }
-\frac{16 \pi ^2 C_A C_F}{3 \epsilon }
\,,
\\ && 
C_{\{1,3,2,0\}}^{\text{div}}=
\log \left(\frac{k^2}{\mu ^2}\right) \left(-\frac{88 C_A C_F}{9 \epsilon }-\frac{8 C_A n_l}{9 \epsilon }+\frac{44 C_A^2}{9 \epsilon }+\frac{16 C_F n_l}{9 \epsilon }\right)
\nonumber \\ && ~~~~~~~~ 
+
\left(\frac{44}{9 \epsilon ^2}+\frac{13 \pi ^2}{18 \epsilon }-\frac{208}{9 \epsilon }\right) C_A C_F
+
\left(-\frac{22}{9 \epsilon ^2}+\frac{8 \pi ^2}{9 \epsilon }+\frac{104}{9 \epsilon }\right) C_A^2
+
\frac{2 \pi ^2 C_F^2}{\epsilon }
\nonumber \\ && ~~~~~~~~ 
+
\left(\frac{4}{9 \epsilon ^2}-\frac{20}{9 \epsilon }\right) C_A n_l
+
\left(\frac{40}{9 \epsilon }-\frac{8}{9 \epsilon ^2}\right) C_F n_l
\,,
\\ && 
C_{\{1,3,0,2\}}^{\text{div}}=
\log \left(\frac{k^2}{\mu ^2}\right) \left(\frac{16 C_A n_l}{9 \epsilon }-\frac{88 C_A^2}{9 \epsilon }\right)
+
\left(\frac{44}{9 \epsilon ^2}-\frac{16 \pi ^2}{9 \epsilon }-\frac{208}{9 \epsilon }\right) C_A^2
\nonumber \\ && ~~~~~~~~ 
+
\left(\frac{40}{9 \epsilon }-\frac{8}{9 \epsilon ^2}\right) C_A n_l
\,,
\\ && 
C_{\{2,3,2,0\}}^{\text{div}}=
-\frac{4 \pi ^2 C_A C_F}{3 \epsilon }
-\frac{2 \pi ^2 C_F^2}{3 \epsilon }
\,.
\eea

The expression of the two-loop Hamiltonian before expansion in $\epsilon$ is fairly lengthy.
We provide the corresponding non-zero $W_{\{i,3,n,2\ell\}}$, defined in eq.~\eqref{defW}, as a list in
a separate file \cite{MathematicaFile} readable, e.g., by {\it Mathematica}.
The list includes the following paramters.
The two-loop counter terms are given by
\bea
&&
\delta _2 Z_2 = \frac{C_F}{512 \pi ^4} \Biggl[\log ^2\left(\frac{\mu ^2}{m^2}\right) \left(-11 C_A+18 C_F+6 n_h+2 n_l\right)+\left(\frac{947}{18}-5 \pi ^2\right) n_h
\nonumber\\&&~~~~~~~~~~~~~~~~~~
+\log \left(\frac{\mu ^2}{m^2}\right) \left(-\frac{215 C_A}{3}+51 C_F+\frac{22 n_h}{3}+\frac{38 n_l}{3}\right)
\nonumber\\&&~~~~~~~~~~~~~~~~~~
+\frac{1}{\epsilon }\left(-\frac{127 C_A}{6}+\log \left(\frac{\mu ^2}{m^2}\right) \left(18 C_F+4 n_h\right)+\frac{51 C_F}{2}+n_h+\frac{11 n_l}{3}\right)
\nonumber\\&&~~~~~~~~~~~~~~~~~~
+\frac{11 C_A+9 C_F-2 n_l}{\epsilon ^2}
+C_A \left(24 \zeta (3)-\frac{1705}{12}+10 \pi ^2-16 \pi ^2 \log (2)\right)
\nonumber\\&&~~~~~~~~~~~~~~~~~~
+C_F \left(-48 \zeta (3)+\frac{433}{4}-\frac{49 \pi ^2}{2}+32 \pi ^2 \log (2)\right)+\left(\frac{113}{6}+\frac{4 \pi ^2}{3}\right) n_l\Biggr] + {\cal O}(\epsilon)
\,,
\nonumber\\&&
\\&&
\delta _2 Z_g = 
\frac{\left(11 C_A-2 n_h-2 n_l\right){}^2}{6144 \pi ^4\epsilon ^2}-\frac{
-5 C_A n_h-5 C_A n_l+17 C_A^2-3 C_F n_h-3 C_F n_l}{1536\pi^4\epsilon }
   \,.
\nonumber\\&&
\eea
The master integrals are given by
\bea
&&
I_{{HH}}^a = \frac{e^{-2 \gamma_E  \epsilon }}{ (4 \pi )^{4-2 \epsilon }} 
\Biggl[
 \frac{3}{2 \epsilon ^2}+\frac{17}{4 \epsilon }+\frac{\pi
   ^2}{4}+\frac{59}{8}
   +\epsilon\left(-\zeta (3)+\frac{65}{16}+\frac{49 \pi
   ^2}{24}\right) 
\nonumber\\&&~~~~~~~~~~~~~~~~~~~~
   +\epsilon ^2
   \left(\frac{151 \zeta (3)}{6}-\frac{1117}{32}+\frac{475 \pi ^2}{48}+\frac{7 \pi
   ^4}{240}-8 \pi ^2 \log (2)\right)
\nonumber\\&&~~~~~~~~~~~~~~~~~~~~
+\epsilon
   ^3 \left(192 \text{Li}_4\left(\frac{1}{2}\right)+\frac{2125 \zeta (3)}{12}-\frac{\pi ^2
   \zeta (3)}{6}-\frac{3 \zeta (5)}{5}-\frac{13783}{64}
\right.
\nonumber\\&&~~~~~~~~~~~~~~~~~~~~ ~~~~~
\left.
   +\frac{3745 \pi ^2}{96}-\frac{103 \pi
   ^4}{96}+8 \log ^4(2)+16 \pi ^2 \log ^2(2)-52 \pi ^2 \log (2)\right)
   \Biggr]
\nonumber\\&&~~~~~~~~~~~~~~~~~~~~
+ {\cal O}(\epsilon^4)
\,,
\\&&
I_{{HH}}^b = \frac{e^{-2 \gamma_E  \epsilon }}{ (4 \pi )^{4-2 \epsilon }} 
\Biggl[
\frac{1}{\epsilon ^2}+\frac{2}{\epsilon }+\frac{11 \pi ^2}{12}-\frac{1}{2}
   +\epsilon 
   \left(\frac{181 \zeta (3)}{12}-\frac{85}{4}+\frac{17 \pi ^2}{24}+\frac{3}{2} \pi ^2 \log
   (2)\right)
\nonumber\\&&~~~~~~~~~~~~~~~~~~~~
   +\epsilon ^2 \left(-36 \text{Li}_4\left(\frac{1}{2}\right)+\frac{157 \zeta
   (3)}{24}-\frac{907}{8}-\frac{373 \pi ^2}{48}+\frac{167 \pi ^4}{72}
\right.
\nonumber\\&&~~~~~~~~~~~~~~~~~~~~ ~~~~~
\left.
   -\frac{3 \log
   ^4(2)}{2}+3 \pi ^2 \log ^2(2)+\frac{3}{4} \pi ^2 \log (2)\right)
\nonumber\\&&~~~~~~~~~~~~~~~~~~~~
+\epsilon ^3 \left(-18 \text{Li}_4\left(\frac{1}{2}\right)+72
   \text{Li}_5\left(\frac{1}{2}\right)-\frac{7733 \zeta (3)}{48}+\frac{2845 \pi ^2 \zeta
   (3)}{72}
\right.
\nonumber\\&&~~~~~~~~~~~~~~~~~~~~ ~~~~~
\left.
   +\frac{15329 \zeta (5)}{40}-\frac{7273}{16}-\frac{4975 \pi ^2}{96}+\frac{107 \pi
   ^4}{90}-\frac{3 \log ^5(2)}{5} -\frac{3 \log ^4(2)}{4}
\right.
\nonumber\\&&~~~~~~~~~~~~~~~~~~~~ ~~~~~
\left.
  +2 \pi ^2 \log ^3(2)+\frac{3}{2} \pi
   ^2 \log ^2(2)+\left(\frac{23 \pi ^4}{5}-\frac{123 \pi ^2}{8}\right) \log
   (2)\right)
   \Biggr]
+ {\cal O}(\epsilon^4)
\,,
\nonumber\\&&~~~~~~~~~~~~~~~~~~~~
\\&&
I_{{HH}}^c = 
-\frac{(4 \pi )^{2 \epsilon -4} \Gamma (3-4 \epsilon ) \Gamma (1-\epsilon )^2 \Gamma (\epsilon ) \Gamma (2 \epsilon -1)}{\Gamma (3-3 \epsilon ) \Gamma (2-2 \epsilon )}
\,,
\\&&
I_{{SS}}^a = -\frac{4^{2 \epsilon -5} \pi ^{2 \epsilon -4} \Gamma (1-2 \epsilon )^2 \Gamma (1-\epsilon ) \Gamma (\epsilon +1) \Gamma (2 \epsilon +1)}{\epsilon ^2 (4 \epsilon -3) (4 \epsilon -1) \Gamma (1-4 \epsilon )}
\,,
\\&&
I_{{SS}}^b = 
 -\frac{16^{\epsilon -2} \pi ^{2 \epsilon -2} \epsilon  \csc (\pi  \epsilon ) \csc (2 \pi  \epsilon ) \Gamma (1-\epsilon )^2}{(2 \epsilon -1) (3 \epsilon -2) (3 \epsilon -1) \Gamma (1-3 \epsilon ) \Gamma (1-2 \epsilon ) \Gamma (\epsilon +1)}
\,,
\\&&
I_{{SS}}^c = 
 \frac{16^{\epsilon -2} \pi ^{2 \epsilon -2} \csc ^2(\pi  \epsilon ) \Gamma (1-\epsilon )^2}{(1-2 \epsilon )^2 \Gamma (1-2 \epsilon )^2}
\,,
\\&&
I_{{SS}}^d = 
 \frac{16^{\epsilon -2} \pi ^{2 \epsilon -4} \Gamma \left(\frac{3}{2}-2 \epsilon \right)^2 \Gamma (1-\epsilon ) \Gamma \left(\epsilon -\frac{1}{2}\right) \Gamma \left(2 \epsilon -\frac{1}{2}\right)}{\Gamma (3-4 \epsilon )}
\,,
\\&&
I_{{SS}}^e = 
 -\frac{8^{2 \epsilon -3} \pi ^{2 \epsilon -2} \csc (\pi  \epsilon ) \Gamma \left(\frac{1}{2}-\epsilon \right)^2 \Gamma \left(\epsilon +\frac{1}{2}\right)}{\Gamma (1-2 \epsilon ) \Gamma \left(\frac{3}{2}-\epsilon \right)}
\,,
\eea
\bea
&&
I_{{SS}}^f = 
\frac{256^{\epsilon -1} \pi ^{2 \epsilon } \Gamma (1-2 \epsilon )^2 \Gamma (2 \epsilon +1)^2}{\Gamma (1-\epsilon )^4 \Gamma (\epsilon +1)^2}
\,,
\\&&
I_{{SS}}^g = 
\frac{ 2^{6 \epsilon -8} \pi ^{2 \epsilon -2} \csc (\pi  \epsilon ) \Gamma \left(\frac{3}{2}-2 \epsilon \right) \Gamma \left(\frac{1}{2}-\epsilon \right) \Gamma \left(2 \epsilon -\frac{1}{2}\right)}{\Gamma (2-3 \epsilon ) \Gamma \left(\frac{3}{2}-\epsilon \right) \Gamma (\epsilon )}
\,,
\\&&
I_{{SS}}^h = 
-\frac{ 4^{3 \epsilon -4} \pi ^{2 \epsilon -\frac{3}{2}} \csc (\pi  \epsilon ) \sec (\pi  \epsilon ) \Gamma \left(\frac{1}{2}-\epsilon \right)^2}{\Gamma \left(\frac{3}{2}-3 \epsilon \right) \Gamma (1-\epsilon )}
\,.
\eea
These results can be found in part in \cite{Piclum:2007an,Schroder:1999sg,Smirnov:2003kc}.

\end{widetext}

\end{document}